\documentclass[aps,prl,reprint,groupedaddress,longbibliography,superscriptaddress]{revtex4-2}
\usepackage{times}
\usepackage{bibentry}
\usepackage{placeins}
\usepackage{amsmath,amsfonts,amssymb}
\usepackage{amsmath}
\usepackage{graphicx}
\usepackage[colorlinks=true, allcolors=blue]{hyperref}
\usepackage{siunitx}
\usepackage[defaultcolor=red]{changes}
\usepackage{soul}
\usepackage{comment}
\usepackage{layouts}
\graphicspath{{../Figures/PDF}}

\DeclareSIUnit \vpp {\ensuremath{\mathrm{V_{pp}}}}
\definechangesauthor[name={Tracy Northup}]{TN}
\definechangesauthor[name={Dmitry Bykov}, color = orange]{DB}

\begin{document}
	\title{A nanoparticle stored with an atomic ion in a linear Paul trap}
\author{Dmitry S. Bykov}
\email[]{dmitry.bykov@uibk.ac.at}
\affiliation{Institut f{\"u}r Experimentalphysik, Universit{\"a}t Innsbruck, Technikerstra\ss e 25, 6020 Innsbruck,
	Austria}
\author{Lorenzo Dania}
\altaffiliation[Present address: ]{Photonics Laboratory, ETH Zürich, CH 8093 Zürich, Switzerland.}
\affiliation{Institut f{\"u}r Experimentalphysik, Universit{\"a}t Innsbruck, Technikerstra\ss e 25, 6020 Innsbruck,
	Austria}
\author{Florian Goschin}
\affiliation{Institut f{\"u}r Experimentalphysik, Universit{\"a}t Innsbruck, Technikerstra\ss e 25, 6020 Innsbruck,
	Austria}
\author{Tracy E. Northup}
\affiliation{Institut f{\"u}r Experimentalphysik, Universit{\"a}t Innsbruck, Technikerstra\ss e 25, 6020 Innsbruck,
	Austria}
\date{\today}

	\begin{abstract}
		Radiofrequency (RF) traps enable highly controlled interactions between charged particles, including reactions between cold molecular ions, sympathetic cooling of one ion species with another, and quantum logic spectroscopy. However, the charge-to-mass ($Q/m$) selectivity of RF traps limits the range of objects that can be confined simultaneously in the same trap. Here, we confine two particles---a nanoparticle and an atomic ion---in the same radiofrequency trap although their charge-to-mass ratios differ by six orders of magnitude. The confinement is enabled by a dual-frequency voltage applied to the trap electrodes. We introduce a robust loading procedure under ultra-high vacuum and characterize the stability of both particles. It is observed that slow-field micromotion, an effect specific to the dual-field setting, plays a crucial role for ion localization. Our results lay the groundwork for controlled interactions between diverse charged particles, regardless of the difference in their charge or mass, with applications from antimatter synthesis to the generation of macroscopic quantum states of motion.		
	\end{abstract}

	\maketitle 
	
The confinement of charged particles in RF traps is a fundamental technique for mass spectrometry~\cite{march2005quadrupole}, cold molecular chemistry~\cite{willitsch2012coulombcrystallised}, and quantum information processing~\cite{wineland1998experimental,leibfried2003quantum,bruzewicz2019trappedion} and has wide-ranging applications from the study of condensed-matter material properties~\cite{nagornykh2017optical} to 
the detection of millicharged dark matter~\cite{budker2022millicharged}.
Typically, a trap is configured to levitate a specific object with a predefined charge-to-mass ($Q/m$) ratio; due to the trap's stability properties, only objects with similar $Q/m$ ratios will be stably confined~\cite{paul1990electromagnetic}.
While this selectivity is advantageous for certain applications, such as mass spectrometry, it limits the range of objects that can be confined simultaneously in the same trap.

To overcome this limitation and trap two species with very different $Q/m$ ratios, a Paul trap driven by two voltage sources has been proposed~\cite{dehmelt1995economic,trypogeorgos2016cotrapping,leefer2016investigation}, targeting antihydrogen synthesis~\cite{dehmelt1995economic,leefer2016investigation} and sympathetic cooling of megadalton particles, including nanoparticles or viruses~\cite{trypogeorgos2016cotrapping}. In this scenario, the frequency and the amplitude of the first voltage source are tuned to provide confinement for the first species, the frequency and amplitude of the second source are tuned for the second species, and due to a careful choice of parameters, the first source does not   compromise the stability of the second species and vice versa. Here, we experimentally demonstrate such a dual-frequency linear Paul trap, which is driven by \SI{17.5}{\mega\hertz} and \SI{7}{\kilo\hertz} voltage sources. This approach allows us to confine vastly different objects, a silica nanoparticle and atomic calcium ions, in the same RF trap, thus creating a hybrid system combining an ultra-high-quality-factor mechanical oscillator~\cite{dania2024ultrahigh} and one of the leading platforms for encoding qubits~\cite{bruzewicz2019trappedion}. The trapped objects differ by six orders of magnitude in their $Q/m$ ratio and by eight orders of magnitude in mass.
	
Let us examine the conditions for stable trapping of a particle---either a nanoparticle or an atomic ion---in the quadrupole electric potential $\Phi$ of a linear Paul trap driven by two voltage sources~\cite{dehmelt1995economic,trypogeorgos2016cotrapping,leefer2016investigation} with frequencies $\Omega_{\text{slow}}$, $\Omega_{\text{fast}}$ and amplitudes $V_{\text{slow}}$, $V_{\text{fast}}$. For the nanoparticle in our system, the main confinement is provided by the slow field, and the influence of the fast field can be neglected. One sees this by comparing the frequencies of the nanoparticle's center-of-mass motion in the separate potentials of the two fields~\footnote{See Supplemental Material at [URL will be inserted by publisher] for stability conditions in a dual-frequency Paul trap, observations of ion loading probabilities and storage times, a calculation of the normal modes, and a discussion of prospects for sympathetic cooling. The Supplemental Material includes Refs.~\cite{bykov2023sympathetic,dania2024ultrahigh,leibfried2003quantum,jin1993precision,hebestreit2018calibration,harlander2011trappedion,brown2011coupled, berkeland98, gulde2001simple}.}:

\begin{equation}\label{eq:sec_fast_nano}
	\frac{\omega_{(\mathrm{n})}|_{\Omega_{\text{fast}}}}{\omega_{(\mathrm{n})}|_{\Omega_{\text{slow}}}} = \frac{\Omega_{\text{slow}}}{\Omega_{\text{fast}}} \frac{V_{\mathrm{fast}}}{V_{\mathrm{slow}}} \approx 10^{-2} ,
\end{equation}
where the approximation in the final step is based on typical voltages used in the experiment (Table~\ref{tab:parameters}).
For the ion, on the other hand, under the pseudopotential approximation and the condition $\Omega_{\text{slow}}\ll\Omega_{\text{fast}}$, the slow field acts as a slowly varying DC offset to the fast field, and the stability condition for the ion motion can be written as~\cite{Note1}
\begin{equation}\label{eq:approx_co-trappin_cond}
	|a_{\mathrm{eff}}|<q_{\mathrm{(i)}}^2/2,
\end{equation}
where $a_{\mathrm{eff}}=4 \kappa \si{\elementarycharge}V_{\text{slow}} / m_{\mathrm{(i)}} r_0^2\Omega^2_{\text{fast}}$, $q_{\mathrm{(i)}}=2\kappa\si{\elementarycharge}V_{\text{fast}} / m_{\mathrm{(i)}}r_0^2\Omega^2_{\text{fast}}$, $r_\mathrm{0}=\SI{0.9}{\milli\meter}$ is the characteristic distance from the electrodes to the trap axis, $\kappa=0.93$ is a geometric factor, $m_{(\mathrm{i})}$ is the mass of the ion, and $e$ is the elementary charge.
Thus, Eq.~\ref{eq:approx_co-trappin_cond} imposes conditions on the fast and slow fields under which co-trapping of a nanoparticle and an ion is possible.

\begin{figure}[t]
	\centering
	\includegraphics[width=0.47\textwidth]{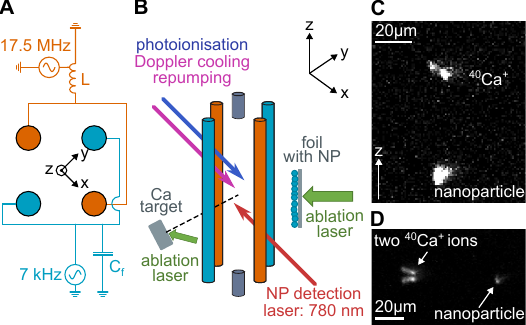}
	\caption{Experimental setup. (a) Dual-frequency drive of the linear Paul trap. (b) Schematic of the procedure for trapping a nanoparticle (NP) and an ion. Along the $z$ axis, DC endcap electrodes are indicated in gray. Not shown: two pairs of compensation electrodes for displacement in the $xy$ plane. (c) Camera image of a $^{40}$Ca$^+$ ion and a nanoparticle confined in the same Paul trap. (d) A second image in which two ions are confined with a nanoparticle.}
	\label{fig:fig_1}
\end{figure}

We implement the dual-frequency trap by applying a voltage oscillating at $\Omega_{\text{fast}} = \SI{17.5}{\mega\hertz}$ to one pair of RF electrodes and a second voltage at $\Omega_{\text{slow}} = \SI{7}{\kilo\hertz}$ to a second pair. 
(A similar configuration is described in Ref.~\cite{foot2018twofrequency} for an experiment with $^{40}$Ca$^+$ ions.) 
The trap schematic is shown in Fig.~\ref{fig:fig_1}a.
The fast voltage signal at $\Omega_{\text{fast}}$ is provided by a low-noise function generator and preamplified by a high-power amplifier. The schematic shows the function generator and preamplifier as a single AC source. We supply the pre-amplified signal to a tap of the inductive coil $L$ of a resonant LC circuit in which the trap electrodes act as a capacitance.
In the case of the slow voltage at $\Omega_{\text{slow}}$, a different function generator drives a low-noise, high-voltage amplifier.
The internal resistance of the slow voltage source and a $\SI{4.7}{\nano\farad}$ capacitor $C_{\text{f}}$ act as a low-pass filter, effectively grounding the fast signal. The slow field sees this capacitor as an open circuit. Similarly, the inductive coil of the resonant circuit acts as a high-pass filter, effectively grounding the slow signal. 
\begin{table}[!h]
	\begin{center}
		\begin{tabular}{c c c c c c} 
			\hline
			& $Q$ ($e$) & $m$ (kg) & $f_{\text{drive}}$ (kHz) & $V_{\text{drive}}$ ($\text{kV}_{\text{pp}}$)& $\frac{\omega}{2\pi}$ (kHz)\\ [0.5ex] 
			\hline\hline
			Ion & 1 & $\SI{6.6e-26}{}$ & \SI{17.5e3}{} & 2.5 & \SI{4e3}{} \\ 
			NP & 800 & $\SI{2e-17}{}$ & 7 & 0.15 & 1.5 \\ 
			\hline
		\end{tabular}
	\end{center}
	\caption{Typical experimental parameters used for co-trapping an ion and a nanoparticle (NP). $Q$ is the charge, $m$ the mass, $f_{\text{drive}}$ the trap-drive frequency, $V_{\text{drive}}$ the trap-drive amplitude, and $\omega$ the frequency of center-of-mass motion in the $xy$ plane of Fig.~\ref{fig:fig_1}.}
	\label{tab:parameters}
\end{table}

As a first step, a nanoparticle is loaded into the trap under ultra-high vacuum via laser-induced acoustic desorption (LIAD)~\cite{AsenbaumKuhnNimmrichterEtAl2013,kuhn2015cavityassisted} combined with temporal control of the Paul-trap potential~\cite{bykov2019direct}. The nanoparticle source is a \SI{300}{\micro\meter} thick aluminum foil with silica nanospheres deposited on the front side; the spheres have a nominal diameter of \SI{300}{\nano\meter}. A pulsed ablation laser (4 mJ, 5 ns) is focused on the back side of the foil (Fig.~\ref{fig:fig_1}b). The geometry of the trap and the source are as described in Ref.~\cite{bykov2019direct}, with the exception that here, the distance between DC endcap electrodes is \SI{3.4}{\milli\meter}. At this stage, only the slow voltage at $\Omega_{\text{slow}}$ is applied to the trap electrodes, with amplitude $V_{\text{slow}}$ = \SI{1.4}{\kilo\vpp}. The endcap voltage is {\SI{56.5}{\volt}.

After loading, we cool the nanoparticle's motion using electrical feedback based on optical detection at \SI{780}{\nano\meter}. There are two differences with respect to the protocol of Ref.~\cite{dania2021optical}: First, the nanoparticle's position along the $x$, $y$, and $z$ axes is detected with a confocal setup implemented with fiber-coupled avalanche photodiodes (APDs)~\cite{vamivakas2007phasesensitive,kuhn2015cavityassisted,xiong2021lensfree,bykov2023sympathetic}. Second, the $x$- and $y$-axis feedback signals are combined and sent to a feedback electrode mounted next to the trap, but the $z$-feedback signal is sent to the lower endcap electrode through a high-pass filter. Feedback cooling operates continuously during the subsequent steps.

Next, we increase the nanoparticle charge $Q_{(\mathrm{n})}$. Recall from Eq.~\ref{eq:approx_co-trappin_cond} that the values of $V_{\text{slow}}$ compatible with co-trapping an ion are bounded from above; a higher value of $Q_{(\mathrm{n})}$ allows the stiffness of the nanoparticle's trap to be maintained while $V_{\text{slow}}$ is reduced~\cite{Note1}. When the ablation laser is directed with a pulse energy of \SI{1}{\milli\joule} to a pure calcium target near the trap ~(Fig.~\ref{fig:fig_1}b), it is observed that $Q_{(\mathrm{n})}$ increases and saturates at around \SI{300}{\elementarycharge} in the absence of a fast field. The presence of a fast field allows even higher values of $Q_{(\mathrm{n})}$ to be obtained during this process.  Thus, at this point we introduce the fast field with $V_{\text{fast}} \approx \SI{1.5}{\kilo\vpp}$. One to three pulses of the ablation laser correspond to an increase of a few elementary charges; we periodically interrupt this stepwise process to reduce $V_{\text{slow}}$ such that the $q_{\text{(n)}}$ parameter remains below 0.9. We stop the charging process at an amplitude $V_{\text{slow}} = \SI{160}{\vpp}$, for which $Q_{(\mathrm{n})} \approx \SI{800}{\elementarycharge}$ is reached. 
Afterwards, both charge and mass are assumed to remain constant: a change in either quantity would shift $\omega_\text{(n)}$, which previously has allowed the addition and subtraction of single elementary charges to be observed \cite{arnold1979determination,schlemmer2001nondestructive,dania2021optical}, but here we continuously monitor $\omega_\text{(n)}$ and find no shift.

Once the nanoparticle has been localized and its charge increased, we  load a $^{40}$Ca$^{+}$ ion into the dual-frequency trap. First, all laser beams required for ion loading, cooling, and detection are aligned to the geometric center of the trap. Here, the trapped nanoparticle serves as a scattering target onto which the attenuated beams are focused. Next, the amplitude of the fast voltage is set to \SI{2.5}{\kilo\vpp}. The voltage on one of the trap's two pairs of compensation electrodes is then increased such that the nanoparticle is displaced in the $xy$ plane, making room for the ion at the trap center. (As the ion's charge is three orders of magnitude smaller than the nanoparticles's charge, its equilibrium position is less affected by the compensation field.) The nanoparticle-detection optics and the feedback-cooling parameters are adjusted to account for this new position. Finally, we operate the ablation laser, still focused on the calcium target, at a pulse energy of \SI{0.5}{\milli\joule} in conjunction with diode lasers at $\SI{375}{\nano\meter}$ and $\SI{423}{\nano\meter}$ for two-step isotope-selective photoionization~\cite{gulde2001simple,hendricks2007alloptical}. The success probability for loading an ion via ablation is similar with and without a nanoparticle~\cite{Note1}. Once an ion is loaded, it is detected and Doppler-cooled via fluorescence on the \SI{397}{\nano\meter} $4^2\text{S}_{1/2} \leftrightarrow 4^2\text{P}_{1/2}$ transition, with repumping on the \SI{866}{\nano\meter} $4^2\text{P}_{1/2} \leftrightarrow 3^2\text{D}_{3/2}$ and \SI{854}{\nano\meter} $4^2\text{P}_{3/2} \leftrightarrow 3^2\text{D}_{5/2}$ transitions.  (Amplified spontaneous emission of the \SI{397}{\nano\meter} diode near the \SI{393}{\nano\meter} $4^2\text{S}_{1/2} \leftrightarrow 4^2\text{P}_{3/2}$ transition occasionally populates the $4^2\text{P}_{3/2}$ state.) All three diode-laser frequencies are locked to a wavemeter.

To verify that a nanoparticle and an atomic ion are co-trapped, we illuminate them with the \SI{397}{\nano\meter}, \SI{866}{\nano\meter}, and \SI{854}{\nano\meter} laser fields and capture their image using an electron multiplying charge-coupled device (EM-CCD) camera. An example image is shown in Fig.~\ref{fig:fig_1}c: nanoparticle and ion are separated by \SI{55\pm10}{\micro\meter}. The laser beams are focused on the ion, and the nanoparticle is in the tail of the Gaussian intensity profile; the camera detects elastically scattered light from the nanoparticle.  Only \SI{397}{\nano\meter} light is used for imaging, and an optical bandpass filter in front of the camera removes other wavelengths. An example of the ion loading process is shown in Supplementary Video 1. The ion typically escapes after several minutes, with the shortest instance being one minute and the longest almost one hour. These storage times are shorter than those for an ion trapped with a single field, which we attribute to the second field, not to the nanoparticle~\cite{Note1}.

	\begin{figure}[t]
	\centering
	\includegraphics[width=0.47\textwidth]{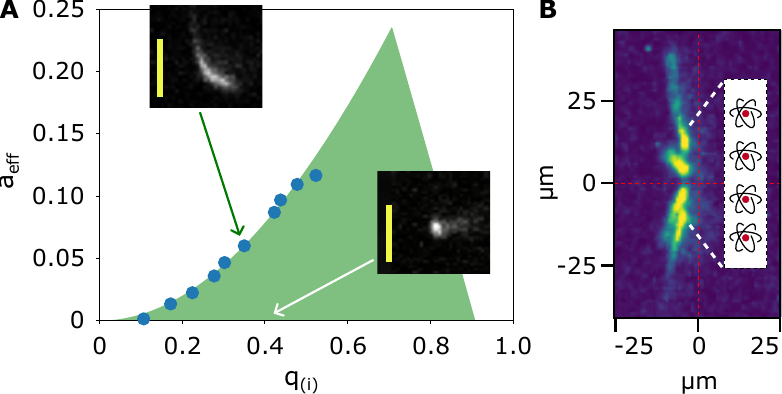}
	\caption{Influence of the slow field on the ion. (a) Blue dots: instability threshold measured with the ion, plotted in terms of the stability parameters $q_\text{(i)}$ and $a_\text{eff}$. Green area: stability diagram, calculated from experimental parameters. Insets are camera images of the ion for different stability regimes; the scale bar is \SI{10}{\micro\meter}. (b) Left: Composite image of the ion at four positions along the $z$ axis for $q_\text{(i)} = 0.4$,  $a_\text{eff} = 0.06$. The positions along the $z$ axis are set by the endcap voltage. Inset: schematic of the ion positions along $z$ axis in the absence of micromotion.}
	\label{fig:fig_2}
\end{figure}
To determine whether the theoretical description above is consistent with our demonstration of co-trapping, we characterize the stability of nanoparticle and ion separately in the dual-frequency Paul trap.
First, a nanoparticle is confined solely with the slow voltage. The fast voltage is then added with increasing amplitude up to \SI{2.5}{\kilo\vpp} while the position and oscillation frequency of the nanoparticle are monitored with a camera and the confocal detection setup, respectively. We find that the nanoparticle remains trapped for all amplitudes and that the presence of the fast voltage increases the nanoparticle's \SI{1}{\kilo\hertz} frequency in the $xy$ plane by \SI{10}{\hertz} for $V_{\text{fast}} = \SI{2.5}{\kilo\vpp}$. These observations are in agreement with Eq.~\ref{eq:sec_fast_nano}.

Next, the ion is confined for several values of $V_{\text{fast}}$. At each setting, $V_{\text{slow}}$ is increased until the ion is expelled from the trap; this threshold amplitude is identified as $V_{\text{slow}}^{\text{max}}$. In Fig.~\ref{fig:fig_2}a, the pair of stability parameters $(q_{\text{(i)}},a_{\text{eff}})$ is plotted for each pair $(V_{\text{fast}},V_{\text{slow}}^{\text{max}})$. Superimposed on these data points is a stability diagram~\cite{leibfried2003quantum}, calculated from the solutions to the Mathieu equation~\cite{Note1}. We see that the experimentally determined thresholds lie at the edge of the stability diagram's stable region, consistent with our expectations. 
The rightmost data point, at $\text{q}_{\text{(i)}}=0.55$, corresponds to $(V_{\text{fast}} = \SI{2.5}{\kilo\vpp},V_{\text{slow}}^{\text{max}}=\SI{260}{\vpp})$.

Camera images provide further evidence of the slow field's influence on ion confinement: the two insets of Fig.~\ref{fig:fig_2}a show that an ion is well localized when the trap is operated deep in the stable region of the stability diagram, but that its position extends over several tens of micrometers at the instability threshold. This elongation is due to micromotion---not the well-known micromotion at $\Omega_{\text{fast}}$ in a single-frequency RF trap~\cite{berkeland98}, but at the second frequency $\Omega_{\text{slow}}$. Figure~\ref{fig:fig_2}b is a composite of four images of this slow-frequency micromotion in the stable region, each obtained for a different position of the ion along the $z$ axis, set by the endcap voltage. We see that the micromotion amplitude increases as the ion is displaced further from the origin. This image underscores the importance of positioning the ion at the slow-micromotion minimum. For future experiments based on the controlled interaction of two particles in a dual-frequency trap, it will be crucial to analyze the impact of slow micromotion. Note that $V_{\text{fast}}$ and $V_{\text{slow}}$ are applied to different electrode pairs, and capacitative coupling could result in different locations for the fast and slow micromotion minima; one could investigate whether combining the voltages on one electrode pair has a different effect on micromotion.
	
\begin{figure}[t]
	\centering
	\includegraphics[width=0.47\textwidth]{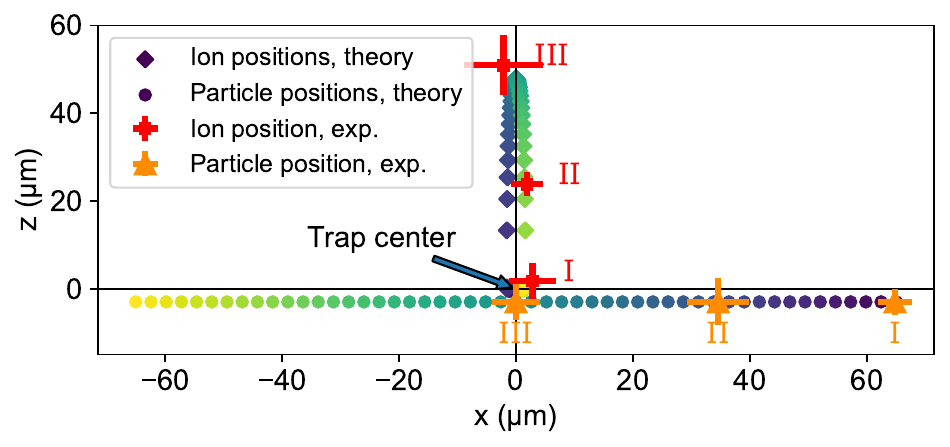}
	\caption{Ion-particle interaction. Circles: fixed nanoparticle positions in the dual-frequency trap. Diamonds: each ion position is calculated from the corresponding nanoparticle position. Squares: ion positions extracted from Fig.~\ref{fig:fig_4}. Triangles: nanoparticle position extracted from Fig.~\ref{fig:fig_4}. The labels next to the data points correspond to the labels in Fig.~\ref{fig:fig_4}c,f. The error bars correspond to the size of the nanoparticle and the ion images in Fig.~\ref{fig:fig_4}. Ion-nanoparticle pairs are indicated with the same color. Calculation parameters: the nanoparticle's charge is \SI{800}{\elementarycharge}, and the ion's secular frequencies along the $x$ and $z$ axes are $\omega_{\text{(i)},x} = \SI{4}{\mega\hertz}$, $\omega_{\text{(i)},z} = \SI{800}{\kilo\hertz}$.
	}
	\label{fig:fig_3}
\end{figure}
Having examined the nanoparticle and ion separately, we turn to the question of where the two particles are located when they are trapped together. Recall, for example, that in the loading procedure, it was necessary to position the nanoparticle such that the ion would be loaded at the trap center, to which the lasers had been aligned. In Fig.~\ref{fig:fig_3}, we calculate the equilibrium position of the ion for different positions of the nanoparticle, using typical experimental parameters. The ion position is determined by the trap potential, Coulomb repulsion from the charged nanoparticle, and DC compensation-electrode voltages $V_{\text{c}1}$ and $V_{\text{c}2}$. Here, fixing the nanoparticle position as an input parameter in the calculation is equivalent to fixing $V_{\text{c}1}$ and $V_{\text{c}2}$. The nanoparticle positions are chosen to lie on the line $(\SI{-65}{\micro\meter} \leq x \leq \SI{-65}{\micro\meter}, y = \SI{0}{\micro\meter}, z = \SI{-3}{\micro\meter})$, where displacement along the $z$ axis is chosen such that the calculated z-distance between the nanoparticle and the ion for the rightmost nanoparticle position "I" matches the experimental distance in Fig.~\ref{fig:fig_4}c-I.
Consider the case of a nanoparticle at $x = \SI{\pm 50}{\micro\meter}$:
the ion is found near the origin, and we infer that the interaction between nanoparticle and ion is negligible compared to the interaction with the trap. As the nanoparticle is brought closer to the origin, the ion is displaced along the $z$ axis, indicating that the particles' interaction has become significant. Thus, this calculation provides intuition for how the competing forces are balanced. It also highlights that there are two ways to trap an ion at the origin, which, as we have just seen, minimizes its micromotion at $\Omega_{\text{slow}}$: the nanoparticle must be displaced either in the $xy$ plane or along the $z$ axis.
	
\begin{figure}[t]
	\centering
	\includegraphics[width=0.47\textwidth]{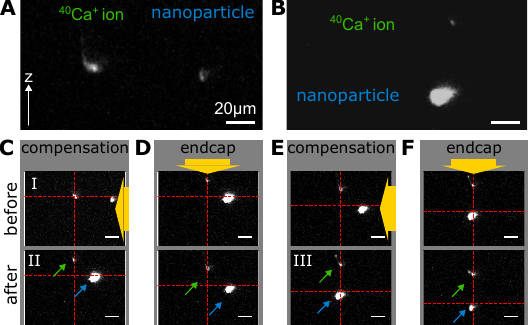}
	\caption{Ion-nanoparticle configurations. The $z$ axis points in the same direction in all images. (a) EM-CCD images of the $xy$ and (b) $z$ configurations. (c) Starting from an $xy$ pair, we decrease the compensation-electrode voltages that have been used to displace the nanoparticle from the trap center. As a result, the nanoparticle shifts towards the center of the trap, and the ion is repelled along the axis of weakest confinement, which is the $z$ axis here. (d) The voltage on one endcap is increased to compensate for this shift; both particles shift away from that endcap. (e) The compensation-electrode voltages are decreased further until the nanoparticle lies on the $z$ axis. (f) The endcap voltage is increased further until the ion is at the original position. The red cross in (c)-(f) indicates the potential minimum of the trap for the ion in the absence of the nanoparticle. The labels I-III correspond to the data-point labels in Fig.~\ref{fig:fig_3}.}
	\label{fig:fig_4}
\end{figure}

Let us refer to the case of nanoparticle displacement in the $xy$ plane as an $xy$ pair and to the case of displacement along the $z$ axis as a $z$ pair. Examples of the two configurations are shown in Figs.~\ref{fig:fig_4}a and \ref{fig:fig_4}b. The nanoparticle in an $xy$ pair experiences excess micromotion and is more sensitive to voltage noise on the trap electrodes than it would be in the $z$ configuration.  Thus, it is desirable to rotate an $xy$ pair into a $z$ pair. A procedure for this transformation is shown in  Figs.~\ref{fig:fig_4}d$-$\ref{fig:fig_4}f, in which the voltages on the compensation electrodes and endcap electrodes are adjusted in four steps. While it is possible to reduce the transformation to two steps, there is a risk that the ion's displacement along the $z$ axis exceeds the \SI{100}{\micro\metre} waist of the Doppler-cooling beams. For the three nanoparticle horizontal positions in Figs.~\ref{fig:fig_4}c$-$\ref{fig:fig_4}f, we have extracted the ion positions relative to the trap center; these were plotted in Fig.~\ref{fig:fig_3}. The extracted ion coordinates are $\left\{[x,z]_{\text{I,II,III}}\right\}_{\text{exp}} =\left\{[3^{+4}_{-4}, 2^{+4}_{-4}], [-2^{+3}_{-3}, 24^{+3}_{-3}], [-2^{+7}_{-7}, 51^{+7}_{-7}]\right\}$. The corresponding theoretically predicted ion positions are $\left\{[x,z]_{\text{I,II,III}}\right\}_{\text{th}} =\left\{[-1^{+0.1}_{-0.1}, 1.9^{+2.6}_{-1.9}], [-1.3^{+0.3}_{-0.6}, 31.0^{+7.0}_{-9.0}], [-0.0^{+0.2}_{-0.2}, 45.9^{+2.5}_{-2.8}]\right\}$, where the uncertainties propagate from the uncertainty of the nanoparticle position. We see good agreement between the theoretically calculated and experimentally measured positions. Following the rotation to a $z$ pair, we observe that if the ion is lost from the trap, another ion can be reloaded directly in this configuration.

We have observed that only a single ion can be loaded in the $z$ configuration, but in the $xy$ configuration, we can co-trap two ions with a nanoparticle, as shown in Fig.~\ref{fig:fig_1}d. To achieve this, the energy of the ablation laser pulse is increased by 10\%, to \SI{0.55}{\milli\joule}; the other co-trapping parameters remain unchanged. We have never observed more than two ions co-trapped with the nanoparticle, but we expect that this should be possible with improvements to the experimental setup. The two-ion limit is not due to the nanoparticle or to the dual-frequency drive: in the absence of the nanoparticle, and with a single frequency $\Omega_{\text{fast}}$ and amplitude $V_{\text{fast}}$, we trap at most two ions. (For lower amplitudes, we are able to trap several ions.)
This points towards problems specific to our trap geometry or RF source: in state-of-the-art traps, it is standard to trap dozens of ions with similar secular frequencies. Co-trapping multiple ions with a nanoparticle is an interesting prospect as one can couple the nanoparticle to different collective motional modes of the trapped ions, which may offer advantages due to noise suppression~\cite{king1998cooling} or due to the mode frequency. 

Calcium ions and a charged silica nanoparticle have been simultaneously confined in a dual-frequency Paul trap. We have studied the ion--nanoparticle pair stability as function of the two RF amplitudes, developed a reliable procedure for loading the trap, and demonstrated a method to swap between different geometric configurations. In-situ charging of the trapped nanoparticle is necessary to reduce the slow-field amplitude to a level that permits ion trapping. The mismatch of six orders of magnitude between the particle's charge-to-mass ratios leads to a differential response to static electric fields, which we found to be beneficial for tuning the ion--nanoparticle separation and maximizing ion-loading efficiencies. As a next step, we plan to analyze the kinetic energies and heating rates of both particles in the trap.

This work has broad applications: it opens up the study of controlled reactions between charged particles with vastly different masses, such as between positrons and antiprotons~\cite{dehmelt1995economic,leefer2016investigation}, or between a wide range of cold molecular ions~\cite{trypogeorgos2016cotrapping}.
Furthermore, our results lay the groundwork for a hybrid system in which a motional state of a nanoparticle is coupled to external or internal degrees of freedom of trapped ions, with prospects for sympathetic cooling~\cite{Note1}. While the center-of-mass motional frequencies of nanoparticles in ion traps lie in the kilohertz regime, ro-vibrational modes have megahertz frequencies~\cite{hoang2016torsional} similar to those of the ions’ center-of-mass motion, enabling both resonant~\cite{chu2017quantum} and dispersive couplings~\cite{arrangoiz2019resolving}. Such a hybrid system has been achieved by coupling single superconducting qubits to a piezoelectric resonator~\cite{oconnell2010quantum}, a surface-acoustic-wave resonator~\cite{manenti2017circuit}, and a bulk-acoustic-wave resonator~\cite{bild2023schroedinger}; the qubit serves to generate and probe quantum states of motion of the resonator~\cite{chu2020perspective}. Building on these pioneering results, levitated particles offer the novel prospect of preparing quantum mechanical superpositions with a spatial extent larger than the mechanical oscillator itself.

Source data are available on Zenodo at \url{https://doi.org/10.5281/zenodo.10716187}

\begin{acknowledgments}
	We thank Giovanni Cerchiari, Carlos Gonzalez-Ballestero, and Elisa Soave for helpful discussions and Rainer Blatt and Roland Wester for critically reviewing the manuscript. This research was supported by the Austrian Science Fund (FWF) under grants I5540, W1259 and Y951.
\end{acknowledgments}

The authors declare no competing interests.

\bibliography{bibliography}
\end{document}


\title{Supplemental material for ``A nanoparticle stored with an atomic ion in a linear Paul trap"}
	\author{Dmitry S. Bykov}
	\email[]{dmitry.bykov@uibk.ac.at}
	\affiliation{Institut f{\"u}r Experimentalphysik, Universit{\"a}t Innsbruck, Technikerstra\ss e 25, 6020 Innsbruck,
		Austria}
	\author{Lorenzo Dania}
	\altaffiliation[Present address: ]{Photonics Laboratory, ETH Zürich, CH 8093 Zürich, Switzerland.}
	\affiliation{Institut f{\"u}r Experimentalphysik, Universit{\"a}t Innsbruck, Technikerstra\ss e 25, 6020 Innsbruck,
		Austria}
	\author{Florian Goschin}
	\affiliation{Institut f{\"u}r Experimentalphysik, Universit{\"a}t Innsbruck, Technikerstra\ss e 25, 6020 Innsbruck,
		Austria}
	\author{Tracy E. Northup}
	\affiliation{Institut f{\"u}r Experimentalphysik, Universit{\"a}t Innsbruck, Technikerstra\ss e 25, 6020 Innsbruck,
		Austria}
	\date{\today}
	\maketitle
	\onecolumngrid
	\appendix
\section{Stability conditions in a dual-frequency Paul trap}
In a coordinate system with the origin at the trap's geometric center, the potential in the $xy$ plane orthogonal to the trap axis can be written as
\begin{equation}
	\Phi(x,y) = \left(V_{\text{slow}}\cos{\Omega_{\text{slow}} t} + V_{\text{fast}}\cos{\Omega_{\text{fast}} t} \right)\frac{x^2-y^2}{2r_0^2}\kappa,
\end{equation}
where $r_\mathrm{0}=\SI{0.9}{\milli\meter}$ is the characteristic distance from the electrodes to the trap axis, $\kappa=0.93$ is a geometric factor, $\Omega_{\text{slow}}$ and $\Omega_{\text{fast}}$ are the two frequencies of the drive voltages, and $V_{\text{slow}}$ and $V_{\text{fast}}$ are their amplitudes. First, we focus on the motion of a nanoparticle along one axis; our goal is to compare the influence of the slow and fast components. The equation of motion along $x$ is
\begin{equation}
	m_{(\mathrm{n})}\ddot{x}=-Q_{(\mathrm{n})}(V_{\mathrm{slow}}\cos{\Omega_{\mathrm{slow}}t}+V_{\mathrm{fast}}\cos{\Omega_{\mathrm{fast}}t})\frac{x}{r_0^2}\kappa,
\end{equation}
where $m_{(\mathrm{n})}$ and $Q_{(\mathrm{n})}$ are the mass and charge of the nanoparticle. If we neglect the fast field and work in the pseudopotential approximation, the oscillation frequency in the effective trap potential can be written as~\cite{berkeland98}
\begin{equation}\label{eq:sec_slow_nano}
	\omega_{(\mathrm{n})}|_{\Omega_{\text{slow}}} = \frac{\Omega_{\text{slow}} q_{(\mathrm{n})}}{2\sqrt{2}},
\end{equation}
where we have introduced the stability parameter 
\begin{equation}\label{eq:qparam_nano}
	q_{(\mathrm{n})} = \frac{2 \kappa Q_{(\mathrm{n})} V_{\mathrm{slow}}}{m_{(\mathrm{n})} r_0^2 \Omega_{\mathrm{slow}}^2}.
\end{equation} 
If, on the other hand, we neglect the slow field, the resonance frequency can be expressed as a fraction of the frequency calculated in Eq.~\ref{eq:sec_slow_nano}:
\begin{equation}\label{eq:sec_fast_nano}
	\omega_{(\mathrm{n})}|_{\Omega_{\text{fast}}} = \frac{\Omega_{\text{slow}}}{\Omega_{\text{fast}}} \frac{V_{\mathrm{fast}}}{V_{\mathrm{slow}}} \omega_{(\mathrm{n})}|_{\Omega_{\text{slow}}} \approx 10^{-2} \omega_{(\mathrm{n})}|_{\Omega_{\text{slow}}},
\end{equation}
where the approximation in the final step is based on typical voltages used in the experiment. Since the stiffness of a harmonic potential is proportional to the square of the frequency, we conclude from Eq.~\ref{eq:sec_fast_nano} that the fast field acts only as a perturbation to the motion of the nanoparticle, which is confined by the slow field, and that no special conditions are imposed on the field parameters.

Next, we turn to the equation of motion of an ion in the dual-frequency field:
\begin{equation}
	m_{(\mathrm{i})}\ddot{x}=-\si{\elementarycharge}(V_{\text{slow}}\cos{\Omega_{\text{slow}} t}+V_{\text{fast}}\cos{\Omega_{\text{fast}}t})\frac{x}{r_0^2}\kappa,
\end{equation}
where $m_{(\mathrm{i})}$ is the mass and $\si{\elementarycharge}$ the elementary charge.
Under the condition $\Omega_{\text{slow}}\ll\Omega_{\text{fast}}$, the slow field acts as a slowly varying DC force, and it is possible to map the ion's equation of motion to the Mathieu equation
\begin{equation}\label{eq:mathieu}
	\ddot{x}+(a_{\mathrm{eff}}+2q_{\mathrm{(i)}}\cos{2t})x=0
\end{equation}
for the stability parameters
\begin{equation}\label{eq:a_eff}
	a_{\mathrm{eff}}=\frac{4 \kappa \si{\elementarycharge}V_{\text{slow}}}{m_{\mathrm{(i)}} r_0^2\Omega^2_{\text{fast}}},
\end{equation} 
\begin{equation}\label{eq:q_fast_ion}
	q_{\mathrm{(i)}}=\frac{2 \kappa \si{\elementarycharge}V_{\text{fast}}}{m_{\mathrm{(i)}}r_0^2\Omega^2_{\text{fast}}}. 
\end{equation}
Here $a_{\mathrm{eff}}$ is proportional to the slow-field amplitude but inversely proportional to the square of the fast-field frequency, in contrast to the typical $a$ parameter for single-frequency Paul traps, in which amplitude and frequency terms refer to the same drive field.
For $q_{\mathrm{(i)}}\ll1$, the stability condition for solutions of the Mathieu equation is approximately~\cite{leibfried2003quantum}
\begin{equation}\label{eq:approx_co-trappin_cond}
	|a_{\mathrm{eff}}|<q_{\mathrm{(i)}}^2/2.
\end{equation}

\section{Ion loading probabilities and storage times}
We begin by discussing the ion loading probabilities. Although we did not collect dedicated data on this parameter, we  summarize our qualitative observations. We use the laser ablation method to load ions into the trap~\cite{gulde2001simple}. Typically, the energy of the ablation pulse is tuned to load 1–2 ions per pulse. The initial adjustment of the pulse energy is carried out at a low RF drive power of 1 W (corresponding to an ion secular frequency of \SI{1}{\mega\hertz}). Here we refer to the fast voltage drive from the main text. We have observed that as the RF drive power is increased, the energy of the ablation pulse must also be increased to maintain a constant loading probability. For example, a tenfold increase in RF power required a 40\% increase in the ablation pulse energy. In contrast, we did not observe any dependence of the loading probability on the presence of the second (slow) field. Finally, we did not observe any change in the loading probability in the presence of the nanoparticle, provided that the Doppler cooling laser (\SI{397}{\nano\meter}) and the repumping lasers (\SI{866}{\nano\meter} and \SI{854}{\nano\meter}) were properly aligned to the ion's equilibrium position, as determined by the trap potential and its interaction with the nanoparticle.

\begin{figure}[h]
	\centering
	\includegraphics[width=0.9\textwidth]{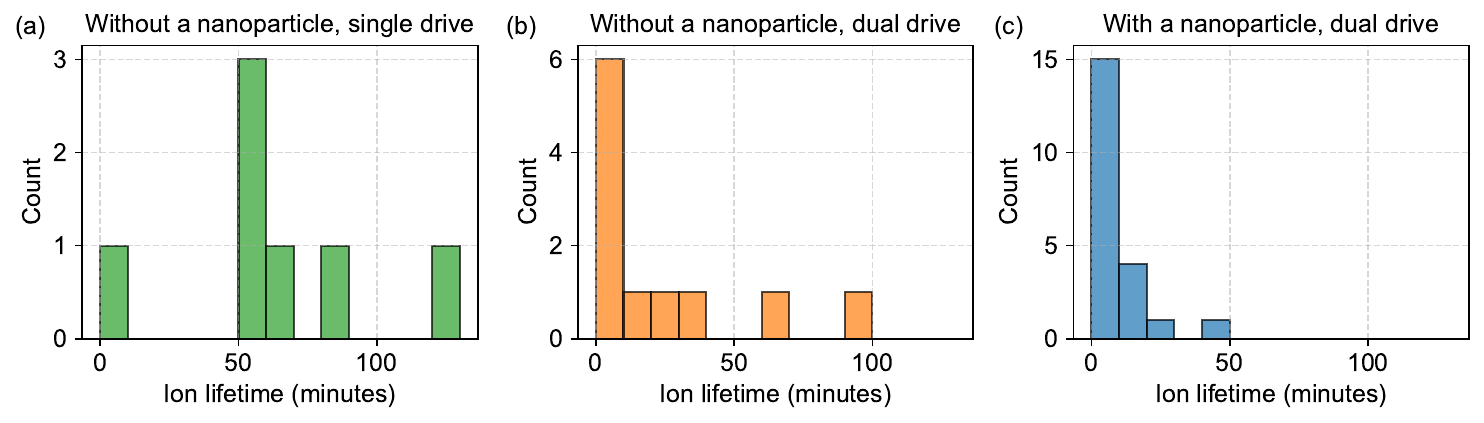}
	\caption{Histograms of ion storage times. (a) The linear Paul trap is driven with a single frequency at high power (\SI{10}-\SI{12}{\watt}). (b) The trap is driven with dual frequencies. (c) The trap is driven with dual frequencies, and a co-trapped nanoparticle is present in the trap.}
	\label{fig:fig_S1}
\end{figure}

We now switch to a discussion of the ion storage time. We have three sets of measurements that allow us to summarize our observations. The first set corresponds to the trap driven solely by the fast drive at high power (\SI{10}-\SI{12}{\watt}), which is the typical power used in our ion-nanoparticle co-trapping experiments. The second set corresponds to the trap driven by both fast and slow drives. In this case, we used voltage settings typical for the co-trapping experiments. The third set corresponds to the trap driven by both drives and a nanoparticle being co-trapped with the ion. These data sets are presented as histograms in Fig.~\ref{fig:fig_S1}. 

Typically, in our setup, we observe ion storage times on the order of a day when the trap is driven with low RF power (\SI{1}{\watt}). Increasing the power reduces these storage times (Fig.~\ref{fig:fig_S1}a). We attribute this to heating of the trap electrodes and subsequent outgassing from the electrodes, increasing the probability of a background-gas collision that converts the atomic ion to a molecular ion. This interpretation is supported by the observation of a slight increase in the ion pump current (\SI{1}-\SI{2}{\nano\ampere}), which indicates a rise in pressure. Our second observation is that adding the second (slow) driving voltage further reduces the ion storage time~(Fig.~\ref{fig:fig_S1}b). Note that, in this case, we have not observed any additional increase in the ion pump current, indicating that the pressure remains stable. Therefore, we do not attribute the decrease in ion storage time to additional heating (and concomitant outgassing) caused by the second voltage. A third and final observation is that, while there is some difference in the storage time distributions for the dual-frequency trap drive with (Fig.~\ref{fig:fig_S1}c) and without (Fig.~\ref{fig:fig_S1}b) a co-trapped nanoparticle, the presence of the nanoparticle does not appear to affect the ion’s storage time significantly. However, making a more definitive claim would require a systematic investigation of ion storage time.

\section{Modes of collective oscillations}
We model an atomic ion and a nanoparticle confined in a dual-frequency Paul trap as two coupled oscillators.
This system has six modes of collective oscillation. Each mode can be expressed as a vector $\left[ q_\mathrm{i}, q_\mathrm{np}, \dot{q_\mathrm{i}},\dot{q_\mathrm{np}}\right]^T = \vec{e}\exp\left(\lambda_q t\right)$, where $q \in \{x,y,z\}$ is a coordinate, ``i'' and ``np'' are labels for the ion and the nanoparticle, $\lambda_q \in \left\{\lambda_{q,\text{in}}, \lambda_{q,\text{out}}\right\}$ is an eigenfrequency, and $\vec{e} \in \left\{ \left[ a_{q,\mathrm{in}}, b_{q,\mathrm{in}}, \lambda_{q,\text{in}}a_{q,\mathrm{in}}, \lambda_{q,\text{in}}b_{q,\mathrm{in}}\right]^T, \left[ a_{q,\mathrm{out}}, b_{q,\mathrm{out}}, \lambda_{q,\text{out}}a_{q,\mathrm{out}}, \lambda_{q,\text{out}}b_{q,\mathrm{out}}\right]^T \right\}$ is an amplitude. The index ``in" corresponds to the mode of collective oscillations in which the ion and the nanoparticle move in phase, while the index ``out" corresponds to the out-of-phase mode. The eigenfrequencies and the amplitudes are defined as the eigenvalues and eigenvectors of matrix~\cite{bykov2023sympathetic}
\begin{equation}\label{eq:matrix}
	\boldsymbol{M}  =
	\begin{bmatrix}
		0 & 0 & 1 & 0\\
		0 & 0 & 0 & 1\\
		-\omega_{0q,\text{i}}^2 + j_{q} & -j_{q} & -\gamma_{q,\text{i}} & 0\\
		-\mu j_{q} & -\omega_{0q,\text{np}}^2 + \mu j_{q} & 0 & -\gamma_{q,\text{np}}
	\end{bmatrix},
\end{equation}
where $\omega_{0q,\text{i}}$ and $\omega_{0q,\text{np}}$ are the resonance frequencies of the ion's and the nanoparticle's center-of-mass (CoM) motion; $\gamma_{q,\text{i}}$ and  $\gamma_{q,\text{np}}$ are total damping rates of the ion's and the nanoparticle's CoM motion; $\mu = m_{\text{i}}/m_{\text{np}}$ is the mass ratio, where $m_\text{i}$ and $m_\text{np}$ are the ion and nanoparticle masses; and $j_{q}$ is a coupling constant. 
As described in the main text, the slow field acts as a slowly varying DC force on the ion, which leads to a modulation of the ion's resonance frequency over time. Here, we consider $\omega_{0q,\text{i}}$ to be the instantaneous frequency at a given point in time.
For axes $x$ and $y$, the coupling constant is $j_{x} = j_{y} = Q_{\text{i}} Q_\text{np} / \left(4\pi\epsilon_0 d^3 m_{\text{i}}\right) = \omega^2_{0z,\mathrm{i}}$; for axis $z$, the coupling constant is $j_{z}=-2Q_{\text{i}} Q_{\text{np}} / 4\pi\epsilon_0 d^3 m_{\text{i}} = -2\omega^2_{0z,\textrm{i}}$, where $Q_{\text{i}}$ and $Q_{\text{np}}$ are the ion's and the nanoparticle's charges and $d$ is the ion-nanoparticle distance. Neglecting the damping, that is, $\gamma_{q,\text{i}} = \gamma_{q,\text{np}} = 0$, we obtain the following expression for the two eigenfrequencies along each spatial direction, where in the final step, we use the fact that the ion is much lighter than the nanoparticle, that is, $\mu \ll 1$:
\begin{equation}
	\begin{split}
		\lambda_{x,\text{in}} &= \sqrt{\frac{j_x +  \mu j_x -\omega_{0x,\textrm{i}}^2 -\omega_{0x,\textrm{np}}^2 + \sqrt{\left(-j_x - \mu j_x+\omega_{0x,\textrm{i}}^2 +\omega_{0x,\textrm{np}}^2\right)^2 - 4\left(-\mu j_x\omega_{0x,\textrm{i}}^2 -j_x \omega_{0x,\textrm{np}}^2 +\omega_{0x,\textrm{i}}^2 \omega_{0x,\textrm{np}}^2\right)}}{2}}\\
		&\stackrel{j_{x}=\omega^2_{0z,\textrm{i}}}{=} \sqrt{\frac{\omega^2_{0z,\textrm{i}} +  \mu \omega^2_{0z,\textrm{i}} -\omega_{0x,\textrm{i}}^2 -\omega_{0x,\textrm{np}}^2 + \sqrt{4\omega_{0x,\textrm{np}}^2 \omega_{0z,\textrm{i}}^2 - 4\omega_{0x,\textrm{i}}^2 \left(\omega_{0x,\textrm{np}}^2 -\mu \omega_{0x,\textrm{i}}^2\right) + \left(\omega_{0x,\textrm{i}}^2 + \omega_{0x,\textrm{np}}^2 - \left(1+\mu\right)\omega_{0z,\textrm{i}}^2\right)^2}}{2}}\\
		&\stackrel{\mu\approx 0}{=} i \sqrt{\omega_{0x,\mathrm{i}}^2 - \omega_{0z,\mathrm{i}}^2}
	\end{split}
	\label{eq:eigen_x}
\end{equation}
\begin{equation}
	\begin{split}
		\lambda_{x,\text{out}} &= \sqrt{\frac{j_x +  \mu j_x -\omega_{0x,\textrm{i}}^2 -\omega_{0x,\textrm{np}}^2 - \sqrt{\left(-j_x - \mu j_x+\omega_{0x,\textrm{i}}^2 +\omega_{0x,\textrm{np}}^2\right)^2 - 4\left(-\mu j_x\omega_{0x,\textrm{i}}^2 -j_x \omega_{0x,\textrm{np}}^2 +\omega_{0x,\textrm{i}}^2 \omega_{0x,\textrm{np}}^2\right)}}{2}}\\
		&\stackrel{j_{x}=\omega^2_{0z,\textrm{i}}}{=} \sqrt{\frac{\omega^2_{0z,\textrm{i}} +  \mu \omega^2_{0z,\textrm{i}} -\omega_{0x,\textrm{i}}^2 -\omega_{0x,\textrm{np}}^2 - \sqrt{4\omega_{0x,\textrm{np}}^2 \omega_{0z,\textrm{i}}^2 - 4\omega_{0x,\textrm{i}}^2 \left(\omega_{0x,\textrm{np}}^2 -\mu \omega_{0x,\textrm{i}}^2\right) + \left(\omega_{0x,\textrm{i}}^2 + \omega_{0x,\textrm{np}}^2 - \left(1+\mu\right)\omega_{0z,\textrm{i}}^2\right)^2}}{2}}\\
		&\stackrel{\mu\approx 0}{=} i \omega_{0x,\mathrm{np}}
	\end{split}
	\label{eq:eigen_x}
\end{equation}
\begin{equation}
	\begin{split}
		\lambda_{y,\text{in}} &= \sqrt{\frac{j_y +  \mu j_y -\omega_{0y,\text{i}}^2 -\omega_{0y,\text{np}}^2 + \sqrt{\left(-j_y - \mu j_y+\omega_{0y,\text{i}}^2 +\omega_{0y,\text{np}}^2\right)^2 - 4\left(-\mu j_y\omega_{0y,\text{i}}^2 - j_y \omega_{0y,\text{np}}^2 +\omega_{0y,\text{i}}^2 \omega_{0y,\text{np}}^2\right)}}{2}}\\
		&\stackrel{j_{x}=\omega^2_{0z,i}}{=} \sqrt{\frac{\omega^2_{0z,i} +  \mu \omega^2_{0z,i} -\omega_{0y,\text{i}}^2 -\omega_{0y,\text{np}}^2 + \sqrt{4\omega_{0y,\text{np}}^2 \omega_{0z,\text{i}}^2 - 4\omega_{0y,\text{i}}^2 \left(\omega_{0y,\text{np}}^2 -\mu \omega_{0y,\text{i}}^2\right) + \left(\omega_{0y,\text{i}}^2 + \omega_{0y,\text{np}}^2 - \left(1+\mu\right)\omega_{0z,\text{i}}^2\right)^2}}{2}}\\
		&\stackrel{\mu\approx 0}{=} i \sqrt{\omega_{0y,\text{i}}^2 - \omega_{0z,\text{i}}^2}
	\end{split}
	\label{eq:eigen_y}
\end{equation}
\begin{equation}
	\begin{split}
		\lambda_{y,\text{out}} &= \sqrt{\frac{j_y +  \mu j_y -\omega_{0y,\text{i}}^2 -\omega_{0y,\text{np}}^2 - \sqrt{\left(-j_y - \mu j_y+\omega_{0y,\text{i}}^2 +\omega_{0y,\text{np}}^2\right)^2 - 4\left(-\mu j_y\omega_{0y,\text{i}}^2 - j_y \omega_{0y,\text{np}}^2 +\omega_{0y,\text{i}}^2 \omega_{0y,\text{np}}^2\right)}}{2}}\\
		&\stackrel{j_{x}=\omega^2_{0z,i}}{=} \sqrt{\frac{\omega^2_{0z,i} +  \mu \omega^2_{0z,i} -\omega_{0y,\text{i}}^2 -\omega_{0y,\text{np}}^2 - \sqrt{4\omega_{0y,\text{np}}^2 \omega_{0z,\text{i}}^2 - 4\omega_{0y,\text{i}}^2 \left(\omega_{0y,\text{np}}^2 -\mu \omega_{0y,\text{i}}^2\right) + \left(\omega_{0y,\text{i}}^2 + \omega_{0y,\text{np}}^2 - \left(1+\mu\right)\omega_{0z,\text{i}}^2\right)^2}}{2}}\\
		&\stackrel{\mu\approx 0}{=} i \omega_{0y,\text{np}}
	\end{split}
	\label{eq:eigen_y}
\end{equation}
\begin{equation}
	\begin{split}
		\lambda_{z,\text{in}} &= \sqrt{\frac{j_z +  \mu j_z -\omega_{0z,\text{i}}^2 -\omega_{0z,\text{np}}^2 - \sqrt{\left(-j_z - \mu j_z+\omega_{0z,\text{i}}^2 +\omega_{0z,\text{np}}^2\right)^2 - 4\left(-\mu j_z\omega_{0z,\text{i}}^2 -j_z \omega_{0z,\text{np}}^2 +\omega_{0z,\text{i}}^2 \omega_{0z,\text{np}}^2\right)}}{2}}\\
		&\stackrel{j_{z}=-2\omega^2_{0z,i}}{=} \sqrt{\frac{-\left(3+2\mu\right)\omega_{0z,\text{i}}^2 -\omega_{0z,\text{np}}^2 - \sqrt{\left(9+4\mu+4\mu^2\right)\omega_{0z,\text{i}}^4 + 2\left(2\mu-3\right)\omega_{0z,\text{i}}^2\omega_{0z,\text{np}}^2 + \omega_{0z,\text{np}}^4}}{2}}\\
		&\stackrel{\mu \approx 0}{=} i \omega_{0z,\text{np}}
	\end{split}
	\label{eq:eigen_z}
\end{equation}
\begin{equation}
	\begin{split}
		\lambda_{z,\text{out}} &= \sqrt{\frac{j_z +  \mu j_z -\omega_{0z,\text{i}}^2 -\omega_{0z,\text{np}}^2 + \sqrt{\left(-j_z - \mu j_z+\omega_{0z,\text{i}}^2 +\omega_{0z,\text{np}}^2\right)^2 - 4\left(-\mu j_z\omega_{0z,\text{i}}^2 -j_z \omega_{0z,\text{np}}^2 +\omega_{0z,\text{i}}^2 \omega_{0z,\text{np}}^2\right)}}{2}}\\
		&\stackrel{j_{z}=-2\omega^2_{0z,i}}{=} \sqrt{\frac{-\left(3+2\mu\right)\omega_{0z,\text{i}}^2 -\omega_{0z,\text{np}}^2 + \sqrt{\left(9+4\mu+4\mu^2\right)\omega_{0z,\text{i}}^4 + 2\left(2\mu-3\right)\omega_{0z,\text{i}}^2\omega_{0z,\text{np}}^2 + \omega_{0z,\text{np}}^4}}{2}}\\
		&\stackrel{\mu \approx 0}{=} i\sqrt{3} \omega_{0z,\text{i}}
	\end{split}
	\label{eq:eigen_z}
\end{equation}
\begin{table}[!h]
	\begin{center}
		\begin{tabular}{|c|c|}
			\hline
			Parameter & Value\\
			\hline
			\hline
			$\omega_{0x,\text{i}}$ & $2\pi\times\SI{4}{\mega\hertz}$\\
			\hline
			$\omega_{0x,\text{np}}$ & $2\pi\times\SI{1.5}{\kilo\hertz}$\\
			\hline
			$\omega_{0z,\text{i}}$ & $2\pi\times\SI{0.8}{\mega\hertz}$\\
			\hline
			$\omega_{0z,\text{np}}$ & $2\pi\times\SI{1}{\kilo\hertz}$\\
			\hline
			$\sqrt{j_x}$ & $2\pi\times\SI{0.8}{\mega\hertz}$\\
			\hline
			$\sqrt{j_z}$ & $2\pi\times\SI{1.1}{\mega\hertz}$\\
			\hline
			$m_{\text{i}}$ & \SI{6.64e-26}{\kilo\gram}\\
			\hline
			$m_{\text{np}}$ & \SI{1.6e-17}{\kilo\gram}\\
			\hline
			$\mu$ & \SI{4e-9}{}\\
			\hline
		\end{tabular}
	\end{center}
	\caption{Typical parameters that characterize the experiment in the main text.}\label{app_table:experimental_parameters}
\end{table}
The general expressions for the eigenvectors are cumbersome. However, using the same approximation for the mass ratio $\mu\ll1$, we obtain the following expressions:
\begin{equation}
		\begin{bmatrix}
			a_{x,\text{in}} \\
			b_{x,\text{in}} \\
			... \\
			...
		\end{bmatrix} = 
		\begin{bmatrix}
			1 \\
			0 \\
			... \\
			...
		\end{bmatrix}
\end{equation}
\begin{equation}
	\begin{bmatrix}
		a_{x,\text{out}} \\
		b_{x,\text{out}} \\
		... \\
		...
	\end{bmatrix} = 
	\begin{bmatrix}
		\omega^2_{0z,\text{i}}/ \left(\omega^2_{0x,\text{i}} - \omega^2_{0z,\text{i}} - \omega^2_{0x,\text{np}}\right) \\
		-1 \\
		... \\
		...
	\end{bmatrix}
\end{equation}
\begin{equation}
	\begin{bmatrix}
		a_{y,\text{in}} \\
		b_{y,\text{in}} \\
		... \\
		...
	\end{bmatrix} = 
	\begin{bmatrix}
		1 \\
		0 \\
		... \\
		...
	\end{bmatrix}
\end{equation}
\begin{equation}
	\begin{bmatrix}
		a_{y,\text{out}} \\
		b_{y,\text{out}} \\
		... \\
		...
	\end{bmatrix} = 
	\begin{bmatrix}
		\omega^2_{0z,\text{i}}/ \left(\omega^2_{0y,\text{i}} - \omega^2_{0z,\text{i}} - \omega^2_{0y,\text{np}}\right) \\
		-1 \\
		... \\
		...
	\end{bmatrix}
\end{equation}
\begin{equation}
	\begin{bmatrix}
		a_{z,\text{in}} \\
		b_{z,\text{in}} \\
		... \\
		...
	\end{bmatrix} = 
	\begin{bmatrix}
		2 \omega^2_{0z,\text{i}}/ \left(3\omega^2_{0z,\text{i}} - \omega^2_{0z,\text{np}}\right)\\
		1 \\
		... \\
		...
	\end{bmatrix}
	\end{equation}
\begin{equation}
	\begin{bmatrix}
		a_{z,\text{out}} \\
		b_{z,\text{out}} \\
		... \\
		...
	\end{bmatrix} = 
	\begin{bmatrix}
		1 \\
		0 \\
		... \\
		...
	\end{bmatrix}
	\end{equation}

Substituting typical parameters of our experiment (Table~\ref{app_table:experimental_parameters}) into Eqs.~\ref{eq:eigen_x}--\ref{eq:eigen_z}, we obtain the eigenfrequencies and eigenvectors listed in Table~\ref{app_table:eigen}. Here we focus on the $x$ and $z$ axes. The analysis for the $y$ axis is analogous to that for the $x$ axis. Additionally, since only two components of the eigenvectors are independent, we list only the amplitudes of motion and omit corresponding time derivatives. The normalized eigenvector of the in-phase $x$ mode indicates that only the ion contributes to the motion. For the out-of-phase $x$ mode, the eigenvector indicates that the oscillation amplitude is primarily determined by the displacement of the nanoparticle, with a minor contribution from the ion. Switching to the $z$ axis, which is aligned with the axis of the ion-nanoparticle Coulomb crystal, we find that the eigenvector of the in-phase mode has similar amplitudes of motion for both ion and nanoparticle. In contrast, the out-of-phase mode involves only the ion's motion.

\begin{table}[!h]
	\begin{center}
		\begin{tabular}{|c|c|c|}
			\hline
			Mode & Eigenfrequency $\lambda_{q,\text{in(out)}}$ & Corresponding eigenvector $\left[ a_{q,\mathrm{in(out)}}, b_{q,\mathrm{in(out)}},..., ...\right]^T$\\
			\hline
			\hline
			$x$, in-phase & $2\pi\times\SI{3.92}{\mega\hertz}$ & $\left[1, 0, ..., ...\right]^T$ \\
			\hline
			$x$, out-of-phase & $2\pi\times\SI{1.5}{\kilo\hertz}$ & $\left[0.04, -1, ..., ...\right]^T$ \\
			\hline
			$z$, in-phase & $2\pi\times\SI{1}{\kilo\hertz}$ & $\left[0.67, 1, ..., ...\right]^T$ \\
			\hline
			$z$, out-of-phase & $2\pi\times\SI{1.38}{\mega\hertz}$ & $\left[1, 0, ..., ...\right]^T$ \\
			\hline
		\end{tabular}
	\end{center}
	\caption{Eigenfrequencies and eigenvectors.}\label{app_table:eigen}
\end{table} 

\section{Sympathetic cooling of the nanoparticle with the ion}
The fact that both the nanoparticle and the ion participate in the $x$ out-of-phase and $z$ in-phase modes suggests that the motion of the nanoparticle could be damped though interaction with the cold ion, i.e., sympathetically cooled. While an experimental demonstration of sympathetic cooling goes beyond the scope of the current work, in this appendix, we theoretically analyse the effect of such cooling on the nanoparticle in our system.

\subsection{Simple estimate of cooling for the center-of-mass mode}

\subsection{Calculation of sympathetic cooling rates} 

Under the influence of random forces, the displacement power spectral densities (PSDs) of the ion and the nanoparticle can be written as
\begin{equation}\label{eq:psd}
	\begin{bmatrix}
		S_{qq,\text{i}}\left(\omega\right) \\
		S_{qq,\text{np}}\left(\omega\right)
	\end{bmatrix}
	=
	\begin{bmatrix}
		\left|\chi_{q,\mathrm{i-i}}\left(\omega\right)\right|^2 & \left|\chi_{q,\mathrm{i-np}}\left(\omega\right)\right|^2\\
		\left|\chi_{q,\mathrm{np-i}}\left(\omega\right)\right|^2 & \left|\chi_{q,\mathrm{np-np}}\left(\omega\right)\right|^2
	\end{bmatrix}
	\begin{bmatrix}
		S^{\text{(rand)}}_{{q,\text{i}}}\left(\omega\right) \\
		S^{\text{(rand)}}_{{q,\text{np}}}\left(\omega\right)
	\end{bmatrix},
\end{equation}
where $\chi_{q,\mathrm{i-i}}, \chi_{q,\mathrm{np-np}}, \chi_{q,\mathrm{i-np}}$ and $\chi_{q,\mathrm{np-i}}$ are response functions and $S_{q,\text{i}}^{\text{(rand)}}$ and $S_{q,\text{np}}^{\text{(rand)}}$ are the PSDs of the random forces acting on the ion and the nanoparticle. The response functions depend on the parameters listed in Table~\ref{app_table:experimental_parameters} and on $\gamma_{q,\text{np}}$ and $\gamma_{q,\text{i}}$. The exact expression can be found in Eqs.~S19--S30 in the supplemental material of Ref.~\cite{bykov2023sympathetic}. We assume that the nanoparticle's damping rate is dominated by the interaction with the background gas and is equal to the value measured in our previous experiment,  $\gamma_{q,\text{np}} = 2\pi\times\SI{69}{\nano\hertz}$~\cite{dania2024ultrahigh}. We assume that the ion's damping rate is due to laser cooling, which we furthermore assume to be Doppler cooling. This can be calculated as~\cite{leibfried2003quantum}
\begin{equation}
	\gamma_{\text{Doppler}} = F_0 \kappa / m_{\text{i}},
\end{equation}
where $F_0$ is defined as
\begin{equation}
	F_0 = \hbar k \Gamma \frac{s/2}{1 + s + \left(2 \Delta / \Gamma\right)^2},
\end{equation}
and $\kappa$ is defined as 
\begin{equation}
	\kappa = \frac{8k \Delta /\Gamma^2}{1 + s + \left(2 \Delta / \Gamma\right)^2}.
\end{equation}
Here, $\hbar$ is the reduced Planck constant; $k = 2\pi / \lambda_{\text{Doppler}}$ is the wave vector of the Doppler-cooling laser; $\Gamma$ is the linewidth of the Doppler-cooling transition; $s=2|\Omega|^2 / \Gamma^2$ is the saturation parameter, which is proportional to the square of the on-resonance Rabi frequency $\Omega$; $\Delta = \omega - \omega_\text{a}$ is the detuning of the Doppler-cooling frequency $\omega = 2\pi c/ \lambda_{\text{Doppler}}$ with respect to the atomic transition frequency $\omega_\text{a}$, and $ c$ is the speed of light. For cooling on the $^{40}$Ca$^+$ transition at $\lambda_{\text{Doppler}} = \SI{397}{\nano\meter}$, the linewidth is $\Gamma~=~1/\SI{7.1}{\per\nano\second}$~\cite{jin1993precision}. 
For a saturation parameter $s=0.5$, we estimate the Doppler cooling rate to be $\gamma_{\text{Doppler}} = 2\pi\times\SI{10}{\kilo\hertz}$. 

According to the fluctuation-dissipation theorem, we can express the PSDs of the random forces acting on the ion and the nanoparticle as 
\begin{align}
	S_{q,\text{i}}^{\text{(rand)}} &= 4 m_{\text{i}} \gamma_{\text{recoil}} k_{\text{B}}T_{\text{phot}} / \pi \label{eq:psd_i},\\
	S_{q,\text{np}}^{\text{(rand)}} &= 4 m_{\text{np}} \gamma_{\text{np,noise}} k_{\text{B}}T_\text{noise} / \pi, \label{eq:psd_np}
\end{align}
where $\gamma_{\text{recoil}}$ is the recoil heating rate of the ion immersed in the bath of Doppler-cooling photons, $\gamma_{\text{np,noise}}$ is the heating rate of the nanoparticle exposed to external noise, $T_{\text{phot}}$ is the temperature of the photon bath, $T_{\text{noise}}$ is the effective temperature of the noise experienced by the nanoparticle, and $k_{\text{B}}$ is the Boltzmann constant. We identify  $\dot{E}_{\text{i}} = \gamma_{\text{np,recoil}} k_{\text{B}}T_\text{phot}$ and $\dot{E}_{\text{np}} = \gamma_{\text{np,noise}} k_{\text{B}}T_\text{noise}$ as the ion and nanoparticle heating rates in units of \SI{}{\joule\per\second}, which allows us to rewrite Eqs.~\ref{eq:psd_i} and \ref{eq:psd_np} as
\begin{align}
	&S_{q,\text{i}}^{\text{(rand)}} = 4 m_{\text{i}}  \dot{E}_{\text{i}} / \pi,\\
	&S_{q,\text{np}}^{\text{(rand)}} = 4 m_{\text{np}} \dot{E}_{\text{np}}  / \pi.
\end{align}
For the nanoparticle heating rate, we use the value measured in our previous experiment, $\dot{E}_{\text{np}} = \SI{2.8e-26}{\joule\per\second}$~\cite{dania2024ultrahigh}. The heating rate of the ion is calculated as~\cite{leibfried2003quantum}
\begin{equation}
	\dot{E}_{\text{i}} = \frac{1}{2m_{\text{i}}}\left(\hbar k\right)^2 \Gamma \left(1 + \xi\right) \frac{s/2}{1 + s + \left(2 \Delta / \Gamma\right)^2},
\end{equation} 
where $\xi$ is a geometrical factor describing the emission and is equal to $2/5$ for dipole radiation~\cite{leibfried2003quantum}. For a saturation parameter of $s=0.5$, we calculate the recoil heating rate of the ion to be $\dot{E}_{\text{i}} = \SI{3.8e-22}{\joule\per\second}$. 

Having defined the response functions and the random noise forces, we can calculate the displacement PSDs of the ion and the nanoparticle and use these PSDs to calculate motional temperatures. Here we focus on the motion of the nanoparticle. Similar to the analysis in Ref.~\cite{bykov2023sympathetic}, we calculate the effective temperature $T_{q,\text{eff}}$ of the nanoparticle's CoM motion based on the mean kinetic energy~\cite{hebestreit2018calibration,bykov2023sympathetic}
\begin{equation}\label{eq:kinetic}
	\langle E_\mathrm{kin} \rangle = \frac{1}{2}m_{\text{np}} \langle \dot{q}^2 \rangle = \frac{1}{2}k_B T_{q,\text{eff}},
\end{equation}
with the variance calculated as $\langle \dot{q}^2 \rangle = \int_{0}^{\infty} S_{\dot{q}\dot{q},\text{np}}(\omega) \,d\omega  = \int_{0}^{\infty}\omega^2 S_{qq,\text{np}}(\omega) \,d\omega$, where $S_{\dot{q}\dot{q}}(\omega)$ is the velocity PSD. We obtain equilibrium temperatures of \SI{2280}{\kelvin} along the $x$ axis and \SI{17}{\kelvin} along the $z$ axis. In the absence of sympathetic cooling, we calculate the equilibrium temperatures of the nanoparticle along the $x$ and $z$ axes to be \SI{4680}{\kelvin}, which is not equal to the room temperature of \SI{300}{\kelvin} because we have set the nanoparticle heating rate to be higher than the damping rate from background gas. This setting reflects the experimental conditions in our system~\cite{dania2024ultrahigh}. 

The parameters and results of the cooling calculation are summarized in Table~\ref{app_table:cooling_parameters}.
\begin{table}[H]
	\begin{center}
		\begin{tabular}{|c|c|}
			\hline
			Input parameter & Value\\
			\hline
			\hline
			$\Gamma$ & $2\pi\times \SI{22.4}{\mega\hertz}$\\
			\hline
			s & 0.5\\
			\hline
			$\lambda_{\text{Doppler}}$ & $\SI{397}{\nano\meter}$\\
			\hline
			$\dot{E}_{\text{np}}$ & \SI{2.8e-26}{\joule\per\second} (radial: \SI{2.8e3}{phonon\,\second^{-1}}, axial: \SI{4.2e3}{phonon\,\second^{-1}})\\
			\hline
			$\dot{E}_{\text{i}}$ & \SI{3.8e-22}{\joule\per\second} (radial: \SI{1.4e5}{phonon\,\second^{-1}}, axial: \SI{7.2e5}{phonon\,\second^{-1}})\\
			\hline
			$\gamma_{\text{Doppler}}$ & $2\pi\times\SI{10}{\kilo\hertz}$\\
			\hline
			$\gamma_{q,\text{np}}$ & $2\pi\times\SI{69}{\nano\hertz}$\\
			\hline
			\hline
			Calculation results & Value\\
			\hline
			$T_{x, \text{eff}}$ & \SI{2280}{\kelvin} (\SI{3.2e10}{phonon})\\
			\hline
			$T_{z, \text{eff}}$ & \SI{17}{\kelvin} (\SI{3.5e8}{phonon})\\
			\hline
		\end{tabular}
	\end{center}
	\caption{Parameters that, together with the parameters from Table~\ref{app_table:experimental_parameters}, were used to calculate the PSDs and CoM temperatures of a nanoparticle sympathetically cooled by a calcium ion.}\label{app_table:cooling_parameters}
\end{table}

In conclusion, calculations using typical parameters for our experiment show that three of the six collective oscillation modes can be used to sympathetically cool the nanoparticle via the Doppler-cooled ion. Along the $z$ axis, where the coupling between the ion's motion and the nanoparticle's motion is strongest, the nanoparticle's temperature can be reduced by two orders of magnitude compared to the case without cooling. In the radial plane, with weaker coupling, the temperature can be reduced by a factor of two. It should be possible to improve the cooling efficiency for motion along the $x$ and $y$ axes by increasing ion's axial confinement relative to its radial confinement. Along the $z$ axis, one could improve the cooling efficiency if the coupling strength could be tuned independently of the ion's bare axial frequency, e.g., by placing the ion and the nanoparticle in separate axial potential wells~\cite{harlander2011trappedion,brown2011coupled}. Finally, using an array of several co-trapped ions would further improve cooling.
\bibliography{bibliography}